\documentstyle[12pt]{article}
\newcommand{\be}{\begin{equation}}
\newcommand{\ee}{\end{equation}}
\newcommand{\bea}{\begin{eqnarray}}
\newcommand{\eea}{\end{eqnarray}}
\begin{document}
\begin{center}

{\large\bf THE SUPERCOMPLEXIFICATIONS AND ODD 
BIHAMILTONIANS STRUCTURES }
\vspace{1.5cm} 

 Z. Popowicz \footnote{E-mail: ziemek@ift.uni.wroc.pl}
\vspace{1.0cm} \\
{Institute of Theoretical Physics, University of Wroc\l aw \\
pl.M.Borna 9, 50-204 Wroclaw, Poland

ziemek@ift.uni.wroc.pl}  
\vspace{1.5cm}
\end{center}

\noindent{\bf Abstract.}
The general method of the complex supersymmetrization (supercomplexifications) of the 
soliton equations with 
the odd (bi) hamiltonian structure is established. New version of the 
supercomplexified  Kadomtsev-Petvishvili hierarchy is 
given. The second odd Hamiltonian operator of the SUSY KdV 
equation generates the odd N=2 SUSY Virasoro - like algebra.
\vspace{2cm}

\section{Introduction}
The Kadomtsev-Petviashvili (KP) hierarchy of integrable soliton nonlinear 
evolution equations [1,2] is among the most important physically relevant 
integrable systems. 
Quite recently a new class of integrable systems motivated by the 
Toda field theory appeared both in the mathematical  and in 
the physical literature .  

On the other side the applications of the supersymmetry (SUSY) to the 
soliton theory provide us a possibility of the  generalization of the 
integrable systems. The supersymmetric integrable equations [3-14] 
have drawn a lot of attention in recent years for a variety of reasons. 
In order to get a supersymmetric theory we 
have to add to a system of k bosonic equations kN fermions and k(N-1) 
boson fields (k=1,2,..N=1,2,..) in such a way that the final theory 
becomes SUSY invariant. Interestingly enough, the supersymmetrizations, 
leads to new effects (not present in the bosonic soliton's theory).  

In this paper we describe a new method of the 
$N=2$ supersymmetrization of the KP hierarchy, first time presented for (M)KdV equation 
in [13]. Our method contains the $N=1$ supersymmetric KdV equation presented by Beckers [7] as a
special case.

Here we would like to present new results on the supercomplexified method.
We  construct the supersymmetric Lax operator for our supercomplex KP hierarchy.
This operator however does not generate the 
supersymmetric conserved currents defined in the whole superspace.
We describe therefore special procedure which allows us to obtain whole 
chain of conserved currents for our supercomplex hierarchy. 

The supercomplex  Korteweg de Vries equation, considered as a special element of our 
KP hierarchy, constitue a
bi-hamiltonian equation and is completely integrable.
The second Hamiltonian operator of the supercomplexified KdV equation generates 
some odd $N=2$ SUSY algebra in contrast to the even algebra SUSY $N=2$ 
Virasoro algebra considered in [8-10]. 

This supercomplexification is a general method. 
In order to obtain the $N=4$ supercomplex version of soliton's equation it is possible to use two 
nonequivalent methods. In both cases we obtain two different generalizations of $N=4$ KdV equation.  
The  model proposed in this paper, differs from the one considered in [9,12].

The idea of introducing an odd hamiltonian 
structure is not new. Leites noticed almost 20 years ago [15], that in the 
superspace one can consider both  even and odd  sympletic structures, with 
even and odd Poisson brackets respectively. The odd brackets (also known as 
antibrackets) have recently drawn some interest in the context of BRST 
formalism in the Lagrangian framework [16], in the supersymmetrical quantum 
mechanics [17], and in the classical mechanics [18,19].

\section{Supercomplexification} 
We shall consider an $N=2$ superspace with the space coordinates $x$ and 
the Grassman coordinates 
$\theta_{1},\theta_{2},\theta_{2}\theta_{1}=-\theta_{1}\theta_{2},
\theta_{1}^{2}=\theta_{2}^{2}=0$. 

The supersymmetric covariant derivatives are defined by 
\be 
\partial = \frac{\partial}{\partial x} ,{~~}{~~} D_{1} = \frac{\partial}{
\partial \theta_{1}} + \theta_{1}\partial , {~~}{~~}
D_{2} = \frac{\partial}{\partial \theta_{2}} + \theta_{2}\partial ,
\ee
with the properties
\be
D_{1}^{2}=D_{2}^{2} = \partial, {~~}{~~} D_{1}D_{2} + D_{2}D_{1} = 0.
\ee
\be
D_{1}^{-1}:= D_{1}\partial^{-1}, {~~} D_{2}^{-1} := D_{2}\partial^{-1}.
\ee
We define the integration over the $N=2$ superspace to be 
\be 
\int dX H(x,\theta_{1},\theta_{2}) = \int dx d\theta_{1}d\theta_{2}
H(x,\theta_{1},\theta_{2}),
\ee
where Berezin's convention are assumed
\be
\int d\theta_{i} \theta_{j} := \delta_{i,j}, {~~} \int d\theta_{i}:=0. 
\ee
We always assume that the components of the superfileds and 
their derivatives vanish rapidly enough. 

Let us now consider some classical evolution equation in the form 
\be
u_{t} = F(u,u_{x},u_{xx},...)= P*grad {~} H(u,u_{x},u_{xx}..),
\ee
where u is considered evolution function, P is the Poisson tensor, H is Hamiltonian of some 
dynamical system and $grad$ denotes the functional gradient.

In order to get the $N=2$ supersymmetric version of the equation (6) let us  
consider the following ansatz which in the next we will call as supercomplexificiation
\be
u=(D_1D_2U) + iU_x,
\ee
where now $U$ is some $N=2$ superfield and $i$ is imaginary quantity $i^{2} = -1$.

Introducing this ansatz to equation (6) we obtain that superfield $U$ evolves in the time as
\be 
U_t = G(U,U_{x},U_{xx},(D_{1}U),(D_{2}U),....) = D_{2}^{-1}D_{1}^{-1}Re(F) = \partial^{-1}Im(F),
\ee
where $Re$ and $Im$ denotes the real and imaginary part respectively.
In order to guarantee validity of (8) we assume that $F$ is chosen in such a way that 
\bea
(D_{1}Re(F)) = (D_{2}Im(F)), \ \\
(D_{2}Re(F)) = -(D_{1}Im(F)),
\eea
always holds. It is a condition of solvability of our construction.

Examples. 

\noindent Let us consider the famous KdV equation 
\be
u_t=-u_{xxx} - 6u_xu.
\ee
The supercomplex version of this equation is
\be
U_t=-U_{xxx} - 6(D_{1}D_{2}U)U_x,
\ee 
and  takes the following form in the components 
\bea
f_t &=& -f_{xxx} + 6gf_x ,\ \\
g_t &=& \partial(-g_{xx} + 3g^2 - 3f_x^2), \ \\
\xi_{1t} &=& -\xi_{1xxx} + 6\xi_{1x}g + 6\xi_{2x}f_x, \ \\
\xi_{2t} &=& -\xi_{2xxx}-6\xi_{1x}f_x + 6\xi_{2x}g,
\eea
where 
\be
U = f +\theta_{1}\xi_{1} + \theta_{2}\xi_{2} + \theta_{2}\theta_{1}g.
\ee
Notice that the bosonic part of the equations (13,14) does not contain any fermions fields. Hence
in some sense this "supersymmetrization is without supersymmetry". 

Now assuming that 
\be
U = \Pi + \theta_{2}\Phi,
\ee
where $\Pi$ and $\Phi$ are $N=1$ superbosonic and superfermionic functions respectively,
our supercomplex KdV equation (12) reduces to 
\bea
\Pi_{t} &=& -\Pi_{xxx} - 6\Pi_x(D_1\Phi), \ \\
\Phi_{t} &=& -\Phi_{xxx} + 6\Pi_x(D_1\Pi_x) - 6\Phi_x(D_1\Phi).
\eea
Notice that when $\Pi =0$ then our equations (20) reduces to the $N=1$ supersymmetric KdV equation 
considered by Beckers in [7]. 

Let us consider the supercomplexified version of the Nonlinear Schr\H odinger equation
as a second example
\bea
f_t &=& -f_{xx} - 2gf^2, \ \\
g_t &=&  g_{xx} + 2fg^2, 
\eea
as the second example. Assuming that 
\bea
f &=& (D_1D_2F) + iF_{x}, \ \\
g &=& (D_1D_2G) + iG_x,
\eea
we obtain 
\bea
F_t &=& -F_{xx} -2\partial^{-1}\Big( 2(D_1D_2 G)(D_1D_2 F)F_x + G_{x}(D_1D_2 F)^{2} - G_{x}F_{x}^{2}\Big), \ \\
G_t &=&  G_{xx} +2\partial^{-1}\Big( 2(D_1D_2 G)(D_1D_2 F)G_x + F_{x}(D_1D_2 G)^{2} - F_{x}G_{x}^{2}\Big).
\eea

\section{N=2 Supercomplexified Kadomtsev-Petviashvili hierarchy} 

It well known that Korteweg de Vries equation as well as the Nonlinear Schr\H odinger equation are the memebrs of 
the KP hierarchy. This hierarchy is defined by the following Lax operator
\be 
Lax := \partial + f\partial^{-1}g,
\ee
which generates the equations by the so called Lax pair representation
\be
\frac{d}{d {} t} Lax = \Big [ Lax ,(Lax)_{+}^{n} \Big ],
\ee
where $n$ is an arbitrary natural number and $(+)$ denotes the projection onto purely superdifferential 
part of  the pseudosuperdifferential element.

In order to define the Lax operator which generate our supercomplexified solitonic equations let us 
first define the supercomplexified algebra of pseudosuperdifferential elements $\Upsilon$  as
a set of the following elements
\be
 \sum_{n=-\infty}^{+\infty}( (D_1D_2F_n) +F_{nx} \partial^{-1} D_1D_2 )\partial^{n},
\ee
where $F_{n}$ is an arbitrary $N=2$ superfield. There are three different projection in this algebra
\be
P_{k}(g) = \sum_{n=k}^{+\infty}( (D_1D_2F_n)+F_{nx} \partial^{-1} D_1D_2) \partial^{n},
\ee
where $g$ is an arbitrary element belonging to $\Upsilon$ while $k$ can take three values $k=0,1,2$ only.

Now in ordedr to obtain the supercomplexified version of the Lax operator (27) it is enought to 
replace $f$ and $g$ by the following substitution
\bea
f \Rightarrow (D_1D_2F) + F_xD_1D_2\partial^{-1}, \ \\
g \Rightarrow (D_1D_2G) + G_xD_1D_2\partial^{-1},
\eea 
in (27) and replace the projection ($+$) by the projection $P_{1}$ defined by (30) with $k=0$ . 

As an example let us consider the Lax pair representation for the supercomplexified N=2 KdV 
equation. The Lax operator for this case is 
\be
Lax:= \partial + \partial^{-1}((D_1D_2F)+ F_xD_1D_2 \partial^{-1}),
\ee
and produces the supercomplexified KdV equation
\be
\frac{d}{d t} Lax :=\Big [ Lax,P_1(Lax^3) \Big ],
\ee
where 
\be
P_1(Lax^3) = \partial^{3} + 3((D_1D_2F) + F_xD_1D_2\partial^{-1})\partial.
\ee

It is interesting to notice that this Lax operator does not produce any conserved currents defined in the 
whole $N=2$ superspace. The traditional supersymmetric residual  definition as the coefficients standing in 
$D_1D_2\partial^{-1}$ in $Lax^n$ gives us that this coefficient after integration over whole superspace is zero.   

\section{Conserved currents for the supercomplexified KdV equation}

It is easy to prove using the symbolic computer computations [20-21] that this equation does not possesses any 
superbosonic currents. On the other side using the same techinque it is easy to find superfermionic 
conserved currents. Let us explain the connections of these currents with the usual (classical) currents 
of the KdV equation. This connection is achieved in four steps. 

First step. Let us supercomplexify an arbitrary conserved current of the classical KdV equation.
By $H_{nr}$ let us denote the real part of n-th conserved current after the supercomplexifications. 

Second step. We compute the usual integral of $H_{nr}$. This can be denoted as
\be
\int H_{nr} dx  = K^{0}_{n} + \int K^{1}_{n} dx.
\ee     
Third step. Now we compute the supersymmetrical integral over first supersymmetrical variable from $K^{1}_{n}$.
It can be symbolicaly denoted as
\be
\int (D_1K^{1}_{n}) dx = H_{1n} + (D_1\int S^{1}_{n}dx).
\ee
Fourth step. Finally we compute the supersymmetrical integral over second supersymmetrical variable which 
is denoted as
\be 
\int (D_2S^{1}_{n}) dx = H_{2n}.
\ee
$H_{1n}$ and $H_{2n}$ are just the conserved superfermionic currents of the supercomplexified KdV equation.

Let us presents  several superfermionic conserved currents of the supercomplexified KdV 
equation
\bea
H_{12}&=& \frac{1}{2} \int dxd\theta_{1}d\theta_{2} {~} U(D_{1}U_x),\ \\
H_{22}&=& \frac{1}{2} \int dxd\theta_{1}d\theta_{2} {~}U(D_{2}U_x), \ \\
H_{13}&=& \frac{1}{2} \int dxd\theta_{1}d\theta_{2} {~}(-(D_2U_{xxx})U + 4(D_2U)(D_1D_2U)U_x), \ \\
H_{23}&=& \frac{1}{2} \int dxd\theta_{1}d\theta_{2} {~}(-(D_1U_{xxx})U + 4(D_1U)(D_1D_2U)U_x). 
\eea
Interestingly these currents does not contain the classical currents of the KdV equation 
in the bosonic or fermionic parts.

We have checked, using the symbolic computations [20-21] thatthis procedure could be aplied to the whole 
supercomplexified KP hierarchy as well.

\section{ Odd Virasoro - like algebra}

It is well known that Korteweg - de Vries equation constitute the so called 
bihamiltonian structure
\be
u_{t}=\partial * grad\Big(\frac{1}{2} \int dx( u_xu_x - 2u^3)\Big ) = Vir * grad\Big (\frac{1}{2} \int dx u^{2}\Big ), 
\ee
where 
\be
Vir := -\partial^{3} -2u\partial -2\partial u.
\ee
The operator $\partial$ and $Vir$ are two Poissons tensors which constitue with two different hamiltonians the 
so called bihamiltonian structure.                                                
Tensor $Vir$ is connected with the Virasoro algebra, realized as the Poison bracket algebra,
through the Fourier decomposition of the field $u$[]. Indeed introducing 
\be
u:=\sum_{k=-\infty}^{+\infty} T_{k} exp(ikx) - \frac{1}{4},
\ee
where $T_{k}$ is some generator, we obtain that 
\be
\Big\{u(x),u(y)\Big\} = Vir * \delta(x-y), 
\ee
are equivalent with the following Virasoro algebra. 
\be
\Big\{ T_{n},T_{m} \Big\} = (n+m)T_{n+m} + \delta_{n+m,0}(n^3-n).
\ee

We successed to find the supercomplexified version of the bihamiltonian structure of the 
supercomplexified KdV equation. Let us define four operators
\bea
P_{21} &:=& 
 D_{1}\partial  -2\partial^{-1}(D_{1}D_{2}U)D_{1} - 2(D_{1}D_{2}U)\partial^{-1}D_{1} \cr
 &&{~~~~~~~~} +  2\partial^{-1}U_{x}D_{2}+2U_{x}\partial^{-1}D_{2}, \ \\
P_{22} &:=& D_{2}\partial  - 2\partial^{-1}(D_{1}D_{2}U)D_{2} -2(D_{1}D_{2}U)\partial^{-1}D_{2} \cr 
 && {~~~~~~~~}- 2\partial^{-1}U_{x}D_{1} - 2U_{x}\partial^{-1}D_{1}, 
\eea
\bea
P_{11} &:=& D_{1}^{-1}, \ \\
P_{12} &:=& D_{2}^{-1}.
\eea
These operators generate the supercomplexified  KdV equation (12)
\be
U_{t} := P_{21} \frac{\delta H_{12}}{\delta U} = P_{22} \frac{\delta H_{22}}{\delta U } 
 =P_{11}\frac{\delta H_{23}}{\delta U}= P_{12}\frac{\delta H_{13}}{\delta U}
\ee
where hamiltonians are defdined by (36-39).

We can construct an $O(2)$ invariant bihamiltonian structure considering 
the linear combination of $ P_{11} \pm P_{12}, P_{21} \pm P_{22} $ with $ 
H_{12} \pm H_{22}, H_{14} \pm H_{24}$ respectively. These structures  
define the same  supercomplexified KdV equation (12). 

Notice that the operators $P_{21},P_{22}$ or $P_{21} \pm P_{22}$ play the 
same role as the Virasoro algebra in the usual KdV equation. There is a 
basic difference - our Hamiltonian operators generates the odd Poisson 
brackets in the odd superspace. In order to obtain the explicit realization 
of this algebra we connect the Hamiltonian operator $P_{21} - P_{22}$ with 
the Poisson bracket 
\be
\{ U(x,\theta_{1},\theta_{2}) , U(y,\theta^{'}_{1},\theta^{'}_{2} \} =
\Big ( P_{21} - P_{22} \Big ) (\theta_{1} - \theta^{'}_{1})
(\theta_{2} - \theta^{'}_{2})\delta(x-y),
\ee
where
\be
U(x,\theta_{1},\theta_{2}) = u_{0} + \theta_{1}\xi_{1} + \theta_{2}\xi_{2} +
\theta_{2}\theta_{1}u_{1}.
\ee
\noindent Introducing the Fourier decomposition of 
$u_{0},\xi_{1},\xi_{2},u_{1}$ 
\be 
\xi_{j} :=\sum^{\infty}_{s=-\infty} G^{j}_{s} e^{isx},  {~~}{~~} j:=1,2,
\ee
\be
u_{0}:=i\sum^{\infty}_{s=-\infty} L_{s} e^{isx}, {~~}{~~} 
u_{1}:=\sum^{\infty}_{s=-\infty} T_{s} e^{isx} - \frac{1}{4},
\ee
in (64)  we obtain
\be
\{ T_{n},T_{m} \} = \{L_{n},L_{m}\} = \{L_{n},T_{m}\} = 0,
\ee
\bea
\{T_{n},G^{i}_{m}\} &=& (n^{2}-1)\delta_{n+m,0} + 
(-1)^{i}2\frac{n-m}{m}T_{n+m}  - 2\frac{n^2-m^2}{m}L_{n+m}, \ \\
\{G^{i}_{n},L_{m}\} &=& (n-\frac{1}{n})\delta_{n+m,0} 
+2\frac{m-n}{nm}T_{n+m} +(-1)^{i}2\frac{m^2-n^2}{nm}L_{n+m},
\eea
\bea
\{G^{i}_{n},G^{i}_{m}\} &=& (-1)^{i}2\frac{m^2-n^2}{nm}\Big( G^{1}_{n+m} +
G^{2}_{n+m}\Big ),\ \\
\{G^{1}_{n},G^{2}_{m}\} &=& -2\frac{m^2-n^2}{nm}\Big(G^{1}_{n+m} -
G^{2}_{n+m}\Big).
\eea
These formulae define the closed algebra with the graded 
Jacobi identity [22]
\be
\sum_{cycl(a,b,c)}(-1)^{[1+a][1+c]}\{a,\{b,c\}\}=0,
\ee
where $[a]$ denotes the parity of $a$. It is the desired odd 
Virasoro - like algebra.

\section{Supercomplexified N=4 KdV equation}
It is possible to obtain in two different manner the supercomplexified  N=4 version of the 
soliton equation. In the first manner we strictly speaking make duble supercomplexifications
of the usual solitons equations while in the second method we supercomplexify the well know
N=2 supersymmetric equations. Let us describe these methods.

In the first method we assume that the superfunction $U$ which satisfy the equation (12) could be 
presented as
\be
U:=(D_{3}D_{4}W) +iW_x.
\ee 
where now $W$ is some $N=4$ bosonic superfunction.
Introducing this ansatz to (12) we obtain that $W$ satisfy the dubly supercomplexified KdV equation
\be
W_{t}:=-W_{xxx} -6\partial^{-1}((D^{4}W)W_{xx}+(D_1D_2W)(D_3D_4W)),
\ee
where $D^{4}=D_{1}D_{2}D_{3}D_{4}$.

For the second method we consider  the $N=2$ supersymmetric KdV equation
\be
\Phi_{t}:=\partial ( -\Phi_{xx} + 3\Phi(D_1D_2\Phi) +\frac{1}{2}(\alpha -1)(D_1D_2\Phi^2)+\alpha \Phi^{3}).
\ee
where $\alpha$ can take three values $-2,1,4$ if we would like to consider the integrable extensions. We assume 
that the superfield $\Phi$ 
satisfy the $N=2,\alpha=4$ SUSY KdV equation (62), takes after the 
supercomplexification the following form
\be
\Phi :=(D_{3}D_{4}\Upsilon) + i\Upsilon_{x},
\ee
\noindent where $\Upsilon$ is some $N=4$ superboson field. 

Substituting this form in (65) we obtain
\bea
\Upsilon_{t} &:=& -\Upsilon_{xxx} + 3(D_{1}D_{2}\left( 
\Upsilon_{x}(D_{3}D_{4}\Upsilon)) \right) -4\Upsilon^{3}_{x} + \cr
&&\Big ( \Upsilon_{x}(D^{4}\Upsilon) + (D_{3}D_{4}\Upsilon)(D_{1}D_{2}\Upsilon_{x}) 
\Big) + 12(D_{3}D_{4}\Upsilon)^{2}\Upsilon_{x}.
\eea

It is the desired 
generalization of the $N=4$ SUSY KdV equation, which is different from the
one considered in [9,12].

Let us remark that usuing the similar methods descrbided earlier it is possible 
to obtain the supercomplexification of the bihamiltonians structures, the Lax pair representations and
the conserved currents. As an example let me present the Lax operator for the equations (64) and (67).

For equation (64) the Lax operator is 
\be
L:=\partial^{2} + (D^{4}W) + (D_1D_2W_x)i_{2} + (D_3D_4W_x)i_1 + W_{xx}i_1i_2
\ee
where $i_1=\partial^{-1}D_1D_2$, $i_2=\partial^{-1}D_3D_4$ and $i_1^{2}=i_{2}^{2}=-1$.

For equation (67) the Lax operator is 
\be
L:=-\Big (D_1D_2 + (D_3D_4\Upsilon)+\Upsilon_x\partial^{-1}D_3D_4\Big )^{2}
\ee

{\bf Acknowledgement} The author wish to thank organizer for 
the scientific discussion and warm hospitability during the conference.
This paper has been supported by the KBN grant 2 P03B 136 16 and by Bogoliubow-Infeld found.


\begin{thebibliography}{99}

\bibitem{1.} Dickey L A, 1991 {\it Soliton Equations and Hamiltonian System} 
(Singapore: World Scientific).

\bibitem{2. } B\l aszak M, 1998 {\it Multi - Hamiltonian Theory of Dynamical 
System} (Berlin: Springer Verlag). 

\bibitem{3 } Kupershmidt B A, Phys.Lett A 109 (1985) 417.

\bibitem{4. } Gervais J L, Phys.Lett B 160 (1985) 125.

\bibitem{5. } Bonora L, Xiong C S, Phys. Lett 317B (1993) 329, Aratyn H, 
Nissimov E, Pacheva S, Vaysburd, Phys.Lett 294 (1992) 167.

\bibitem{6. } Kulish P, Lett. Math. Phys. 10 (1985) 87.

\bibitem{7.}Becker.K , Becker M Mod.Phys.Lett A 8 (1993) 1205.

\bibitem{8.} Laberge C A, Mathieu P, Phys.Lett B 215 (1988) 718.
\bibitem{9.} Chaichian M, Lukierski J, Phys.Lett B 183 (1987) 169.
\bibitem{10.} Oevel W, Popowicz Z, Commun.Math.Phys. 136 (1991) 441.
\bibitem{11.} Popowicz Z, Phys.Lett. A 174 (1993) 411.
\bibitem{12.} Delduc F, Ivanov E, Phys.Lett B 309 (1993) 312.
\bibitem{13.} Popowicz Z, Phys.Lett B 459 (1999) 150.
\bibitem{14.} Dagris P, Mathieu P, Phys.Lett A 176 (1993) 67.
\bibitem{15.} Leites D A, Dokl.Akad.Nauk SSSR  236 (1977) 804 (in Russian).
\bibitem{16.} Batalin A, Vilkovsky G A, Phys. Lett B 102 (1981) 27,
Nucl.Phys. B234 (1984) 106.
\bibitem{17.} Volkov D A, Soroka V A, Pashnev A I, Tkach V I, JETP Lett.
44 (1986) 55 (in Russian).
\bibitem{18.} Kupershmidt B A, Lett.Math.Phys. 9 (1985) 323, Frydryszak A,
J.Phys.A 26 (1993) 7227, Frydryszak A, Lett.Math.Phys. 44 (1998) 89,
Khudaverdian, J.Math.Phys. 32 (1991) 1934.
\bibitem{19.} Soroka V A, Phys.LettB 451 (1999) 349 ,{\it Degenerate Odd 
Poisson Bracket on Grassmann Variables} hep-th 9811223.
\bibitem{20.} Hearn A, 1995 {\it REDUCE Users's Manual version 3.6}
\bibitem{21.} Popowicz Z, {\it SUSY2}  Comput.Phys.Commun 100 (1997) 277.
\bibitem{22.} \L opuszanski J {\it An Introduction to Symmetry and 
Supersymmetry in Quantum Field Theory} (World Scientific) 1991,
Kupershmidt B, 1987 {\it Elements of Superintegrable 
Systems} (Dordecht: Kluwer).

\end{thebibliography}
 \end{document}